\documentclass[a4paper,twoside,12pt]{revtex4}
\usepackage{epsfig}
\usepackage{graphicx}
\begin{document}
\title{Coexisting charge modulation and ferromagnetism produces long
  period phases in manganites: new example of electronic soft matter.}
\author{G.C. Milward, M.J. Calder\'on, P.B. Littlewood}
\affiliation{
Cavendish Laboratory, Cambridge University, Madingley Road,
Cambridge CB3 0HE, UK}
\date{\today}

\maketitle

{\bf
The phenomenon of colossal magnetoresistance in manganites is generally
agreed to be a result of competition between crystal phases with different
electronic, magnetic, and structural order; a competition which can be
strong enough to cause phase separation between metallic ferromagnet and
insulating charge modulated states
\cite{reva,mathur-littlewood1,mathur-littlewood2,dagottobook,uehara99}.
Nevertheless, closer inspection of
phase diagrams in many manganites reveals complex phases where the two order
parameters of magnetism and charge modulation unexpectedly coexist\cite{loudon02,chen96}.
Here we show that such experiments can be naturally explained within a 
phenomenological Ginzburg-Landau theory. 
In contrast to models where phase separation originates from disorder
  \cite{moreo00} or as a strain induced kinetic
phenomenon \cite{ahn04}, 
 we argue that magnetic and
  charge modulation coexist in new thermodynamic phases.
This leads to a rich diagram of equilibrium phases,
qualitatively similar to those seen in experiment.
The success of this model argues for a fundamental reinterpretation
of the nature of charge modulation in these materials from a localised
to a more
extended "charge density wave" picture.
The same symmetry
considerations that favour textured coexistance of charge and magnetic
order
may apply to many electronic systems with competing
phases. The resulting  "Electronically soft" phases of matter with
incommensurate, inhomogeneous and mixed order may be general phenomena
in correlated systems.
}\\
\\


The manganese perovskites
($RE_{1-x}^{3+}AE^{2+}_xMnO_3$, RE rare earth, AE alkaline earth)
provide a laboratory to study the interplay of a variety of magnetic,
electronic and structural phases of matter in a strongly correlated
electronic system. As in many strongly correlated electronic systems, the
basic paradigm for manganite physics is the competition between the
delocalising effects of the electron kinetic energy and the localising
effects of the Coulomb repulsion, aided by coupling to lattice degrees of
freedom. When the kinetic energy is dominant, one finds a metallic ground
state with ferromagnetic alignment of the core moments. When the localising
effects preponderate, instead we see charge and/or orbitally ordered ground
states with substantial local lattice distortions from the near cubic
symmetry of the metal, along with insulating behaviour and
antiferromagnetism. One may tune between these two phases by many external
parameters, especially chemical substitution, but also lattice strain, and
magnetic field. The competition between metal and insulator is famously
evident in the phenomenon of bulk colossal magnetoresistance, where a
magnetic field tunes the conductivity of the material, and even more clearly
in the strong tendency toward phase separation and inhomogeneity and regimes
of percolative transport. 


The origin of charge and orbital ordered phases is still the subject of debate.
Charge modulation has been
traditionally seen as the ordering of Mn$^{4+}$ and Mn$^{3+}$
ions\cite{goodenough55}. More recently, the charge disproportionation of
the Mn ions has been argued to be much smaller than one\cite{garcia01,brey04,coey04}
but still the idea of
two kinds of cation forming stripes prevails and is used to
interpret the experiments.
In such a scenario, one expects a density $x$
of one kind of cation and $1-x$ of the other.
The charge modulation would be given by the averaged wave-vector
$\mathbf {q} \approx (1-x) \mathbf {a}^*$  with $\mathbf a^*$ the
  reciprocal lattice vector, aside from possible commensuration
  effects near special dopings
($x=1/2,2/3,3/4$).

\begin{figure}
\epsfig{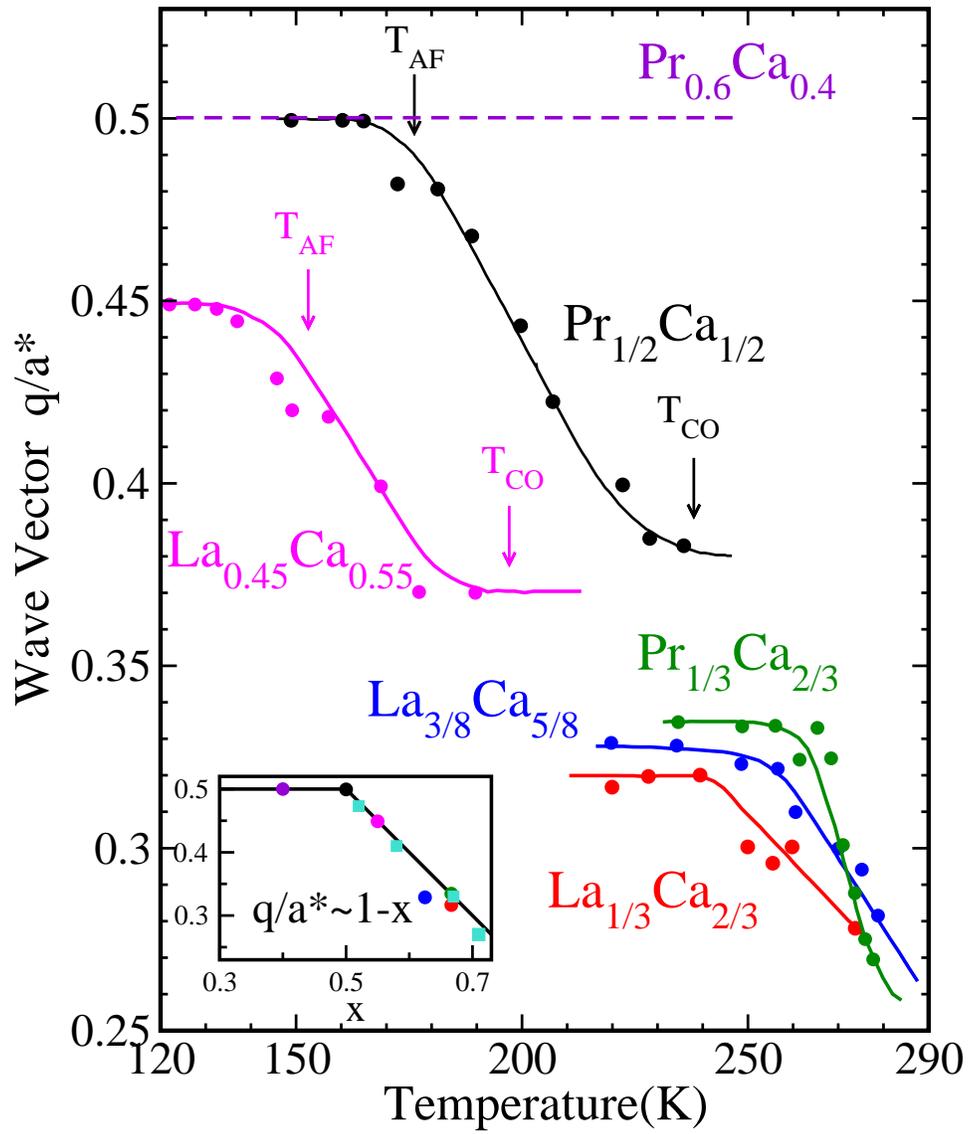}
\caption{Wave vector of the modulation $q/a*$ versus temperature for
  Pr$_{1-x}$Ca$_x$MnO$_3$ and La$_{1-x}$Ca$_x$MnO$_3$ at different
  dopings ($x \ge 0.5$ data taken from Figs. 1 and 2 in
  \cite{chen99} and $x=0.4$ taken from \cite{zimmermann01}). The same
  kind of behaviour has been reported for La$_{1/2}$Ca$_{1/2}$MnO$_3$
  \cite{chen96} (not shown here).
  At low temperatures $q/a* \approx (1-x)$, as shown in the
  inset, and decreases above the N\'eel temperature $T_{AF}$.
  The inset shows $q/a*$ versus $x$ as in
  \cite{chen97,larochelle01}.  The circles
  correspond to the low temperature values of the curves in the main
  panel while the squares are data for La$_{1-x}$Ca$_x$MnO$_3$ taken
  from \cite{loudon04}.
}
\label{fig:exp}
\end{figure}

However, this picture is not compatible with the experimental findings on
commensurate and incommensurate modulation, summarised in Fig.\ref{fig:exp}.
 At half-doping, commensurate
modulation is expected and found at low temperatures, together with
antiferromagnetism of the CE type \cite{goodenough55}.
Above the N\'eel temperature, though, the modulation is
incommensurate \cite{chen96,mori99,kajimoto01}. When found to be
charge modulated, underdoped samples
($x<0.5$) do not show the
relation  $\mathbf q \approx (1-x) \mathbf a^*$,  rather the
modulation wave-vector is always commensurate with $\mathbf q=0.5 \,
\mathbf a^*$
\cite{jirak85,zimmermann01,larochelle01,chen97} and independent of temperature.
 Finally, the overdoped
samples ($x>0.5$) show the expected incommensurate
wave-vector~\cite{chen97} below the
N\'eel temperature, decreasing above it \cite{chen99}. However, no
sign of discommensurations, but a uniform incommensurate modulation,
has been found in recent experiments on La$_{1-x}$Ca$_{x}$MnO$_3$
\cite{loudon04}.

Another experimental conundrum is the coexistence of
charge modulation and ferromagnetism despite their natural antipathy.
For instance, at half-doping, the incommensurate modulation above the N\'eel
temperature
is accompanied by ferromagnetism
\cite{chen96}. A different electronic phase showing ferromagnetism and charge
modulation has also been found at low temperature in 
La$_{0.5}$Ca$_{0.5}$MnO$_3$ \cite{loudon02}.
Slightly overdoped samples can also be ferromagnetic above N\'eel temperature
 \cite{schiffer,kajimoto99}.
Another example of coexistence is given by the underdoped ($0.3<x<0.5$)
Pr$_{1-x}$Ca$_{x}$MnO$_3$. The ground state is commensurate and
charge-modulated
\cite{zimmermann01,larochelle01}  but the
antiferromagnetism is canted \cite{jirak85,tomioka96,yoshizawa95},
showing a
ferromagnetic component coexisting with the commensurate charge modulation.

One might propose to explain these phenomena in terms of microscopic
theory incorporating the many different couplings and microscopic
degrees of freedom, but this is a daunting task that might not be
illuminating owing to its complexity.
Here we propose a simple and more transparent phenomenological
approach to this problem by means of Ginzburg-Landau theory.
We will show how the close interplay of
ferromagnetism and charge modulation conspires to reproduce the
experimental findings just discussed.

Ginzburg-Landau theory allows the study of phase transitions in a
phenomenological way and it consists in expressing the free energy as
a power expansion of the order parameters and their gradients.
The order parameters we consider here are the
magnetisation $M(r)$ and the charge-orbital modulation
 $\psi(r)=\rho(r) e^{i \left(
  {\mathbf {Q_c}} . \mathbf r+\phi(r) \right)}$. $r$
is the spatial coordinate, $\rho$ is the amplitude of the modulation,
${\mathbf {Q_c}}={{\mathbf a^*}\over{n}}$ is a wave-vector
commensurate with the
lattice  and $\phi$
is the phase that would incorporate structures with incommensurate
periodicities. To simplify the discussion we study a one dimensional
scalar modulation, since charge modulation within a domain occurs only
in one direction. $n=4$ gives the correct periodicity for
the lattice distortions measured in
$x=0.5$ as,  though the charge modulation has period $2$, the orbital order
follows a zig-zag pattern with period $4$.
Notice that
if $\nabla \phi=0$, $\psi$ is a wave of amplitude $\rho$ and
wave-vector commensurate with the lattice.
If $\nabla \phi \ne 0$, the wave-vector is ${\mathbf
    {Q_c}}+\langle \nabla \phi \rangle$ and therefore, in general, cannot
be expressed as a simple rational number of $\mathbf a^*$.

The free energy density can be separated into three contributions:
 magnetisation, charge
 modulation and coupling terms. The first two are
\begin{equation}
 {\cal F}_M={{1}\over{2}}a_M(T-T_c) M^2+
{{1}\over{4}}b_M  M^4 +{{1}\over{2}}\xi_M^2
 (\nabla M)^2,
\end{equation}
\begin{equation}
{ \cal F}_{\psi}={{1}\over{2}} a_{\rho}(T-T_{CO})\rho^2+{{1}\over{4}} b_{\rho}
 \rho^4
+{{1}\over{2}}\xi_{\rho}^2(\nabla \rho)^2+{{1}\over{2}} \xi_{\rho}^2 \rho^2
 \left( \nabla\phi  -q_o \right)^2
+{{1}\over{n}} \eta \rho^n \cos(n\phi)
\label{eq:f-phi}
\end{equation}

The magnetic energy, $ {\cal F}_M$,
 taken alone will describe a phase transition to
homogenous magnetism below the Curie temperature $T_C$.
${\cal F}_{\psi}$ is the free energy extensively used to study
commensurate-incommensurate phase transitions of charge density waves,
spin density waves or modulated lattice distortions \cite{toledano}.
$q_o=1/2-x$ is the predicted deviation (by chemical composition)
from commensurability around $x=0.5$.
The term ${{1}\over{2}} \xi_{\rho}^2 \rho^2 \left( \nabla \phi  -q_o \right)^2$
favours a uniform incommensurate modulation with
$\nabla \phi=q_o$. On the other hand
${{1}\over{n}} \eta \rho^n \cos(n\phi)$
is an Umklapp term that favours commensurability with $\phi=2 \pi
j/n$, $j$ integer
(for $\eta<0$). Taken alone, this describes two phases. Upon cooling
below $T_{CO}$, the amplitude $\rho$ of the charge density wave rises from zero
 but provided $n>2$, the Umklapp term is small and the modulation
is incommensurate. As temperature is lowered,
$\rho$ grows, the Umklapp term may become dominant and a lock-in
transition occurs if $\eta$ is comparable to $\xi_{\rho}^2$.

We now discuss coupling between the two order parameters. The lowest order
coupling term which arises is $d_1\rho^2M^2$ so that there is a free energy
penalty for
homogeneous coexistence of magnetism and charge modulation. Were this the only
coupling term the free energy would be generally stabilised either by a
homogenous magnetisation or by charge modulation, depending on which transition
temperature is the larger. Next one can of course introduce uniform coupling
terms of higher powers of $M$ and $\rho$, but they make no qualitative
changes unless they have a negative sign.  More interesting is
that there is a leading order coupling term in the {\em gradient} of the
form $d_2 \rho^2 M^2 (\nabla \phi-q_o)$.
The fact there is a term {\em linear} in the
gradient is expected because there is no symmetry about $x=1/2$; different
signs of the gradient correspond physically to compression or extension of
the CDW period, i.e. to extra  "3+" or "4+" sites. One can also
justify this term microscopically:
if we consider the effect of charge modulation on the Fermi
surface, then it is clear that if we choose a wave vector which does not match
the chemical doping, one will be left with small pockets of carriers at the
Fermi surface; these metallic electrons (or holes)
are then available to mediate double
exchange and thereby promote ferromagnetism. The asymmetry around
$x=1/2$ is due to the asymmetry between electron and hole pockets.
Now note that this
gradient term can be incorporated into Eq. \ref{eq:f-phi} by completing
the square, and replacing $q_o$ by

\begin{equation}
q_{eff} = q_o - \frac{d_2}{\xi_{\rho}^2} M^2 = \frac{1}{2} - x -
\frac{d_2}{\xi_{\rho}^2} M^2  \;\; .
\label{eq:qeff}
\end{equation}

The sign of $d_2$ is unknown {\it a priori} and we here choose it to be
positive. Once this sign is fixed, however, this gradient coupling has
profound consequences for the phase diagram. First, note that even if we are
at commensurability ($x=1/2$), if magnetism is present, then there is a
tendency to {\em incommensurate} charge modulation. This reproduces the
experiments of Chen et al. \cite{chen96}, on La$_{1/2}$Ca$_{1/2}$MnO$_3$,
where the onset of charge modulation is incommensurate, and accompanied by
ferromagnetism - which is replaced by N\'eel order at the transition to the
commensurate phase. The incommensurate phase of
Pr$_{1/2}$Ca$_{1/2}$MnO$_3$ (see Fig. \ref{fig:exp}) is paramagnetic
\cite{mori99} but accompanied by the onset of ferromagnetic spin
fluctuations \cite{kajimoto01,kajimoto98}.
The second feature of this term is that if $x<1/2$, it
is possible for coexisting magnetism to ``cancel'' the chemical tendency to
incommensurability, and we note that canted magnetism is generally reported
\cite{jirak85,tomioka96,yoshizawa95} in the underdoped manganites that show
 commensurate charge modulation. No
such cancellation is possible for $x>1/2$, and thus the dog-leg dependence
of wave-vector on doping shown in the inset to Fig. \ref{fig:exp} is indicated.

\begin{figure}
  \epsfig{file=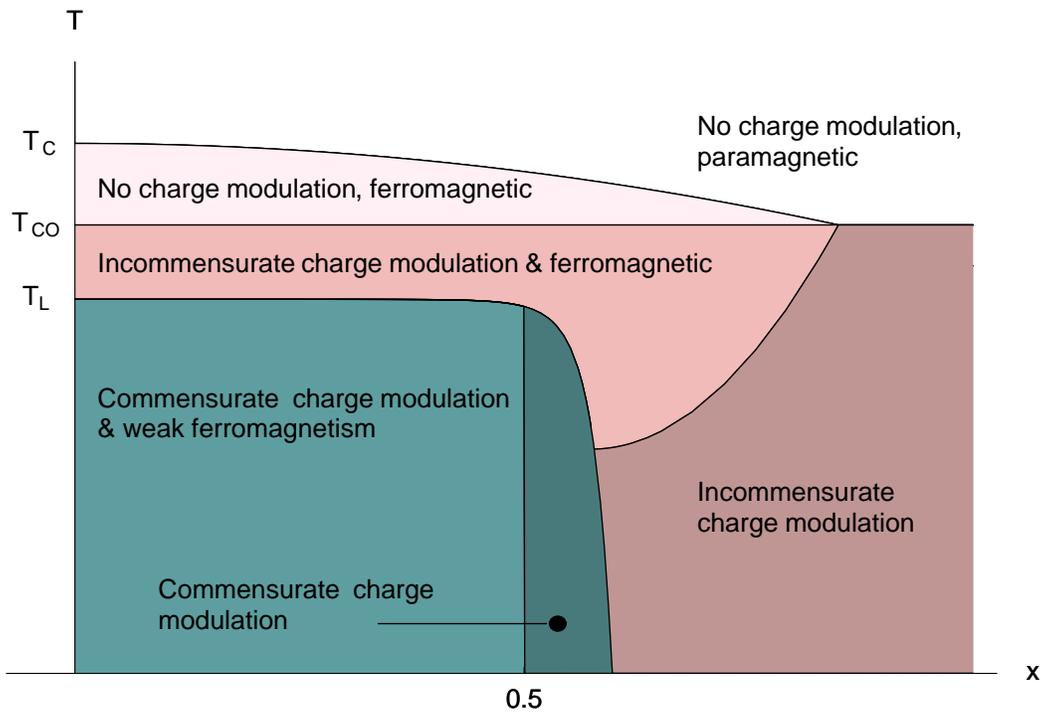, width=6in}
\caption{Schematic phase diagram which results from the minimisation of the free
  energy. The scale of the axes depends on the particular
  parameters used.
  The commensurate order phase just above $x=0.5$ has
  not been observed but we predict it can be relevant for dopings very
  close to $x=0.5$ for highly
  insulating manganites.
  The complex phases arise provided $T_C
  >T_{CO}$, a condition that can be relaxed if the model is extended
  to account for discontinuous phase transitions.
  The values of $T_{CO}$ and $T_C$ are direct parameters in the model
  whereas the lock-in temperature $T_L$ is a consequence of the competition between
  the Umklapp and incommensurate modulation terms.
}
\label{fig:GLphasediagram}
\end{figure}

\begin{figure}
  \epsfig{file=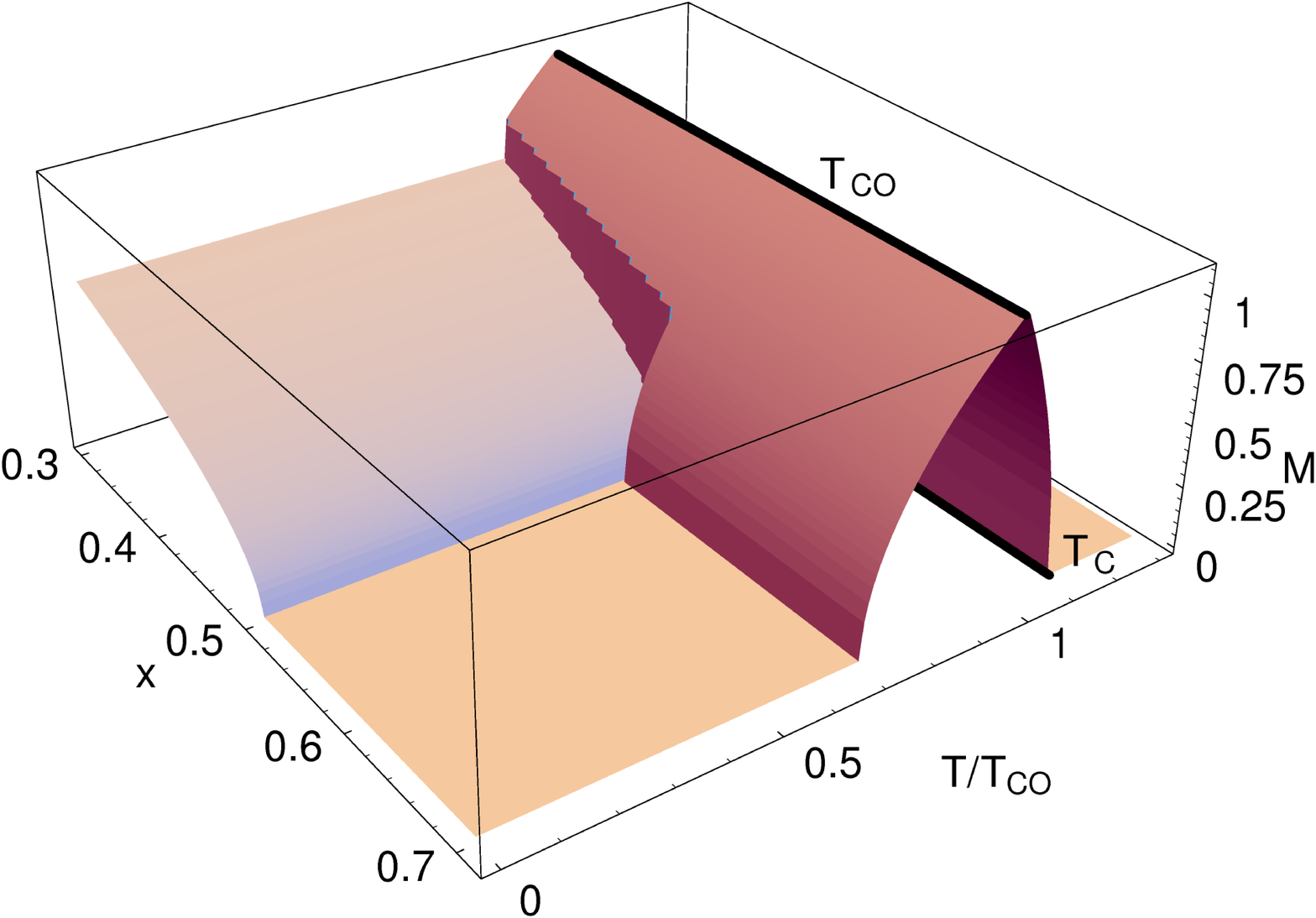, width=3.5in}
  \epsfig{file=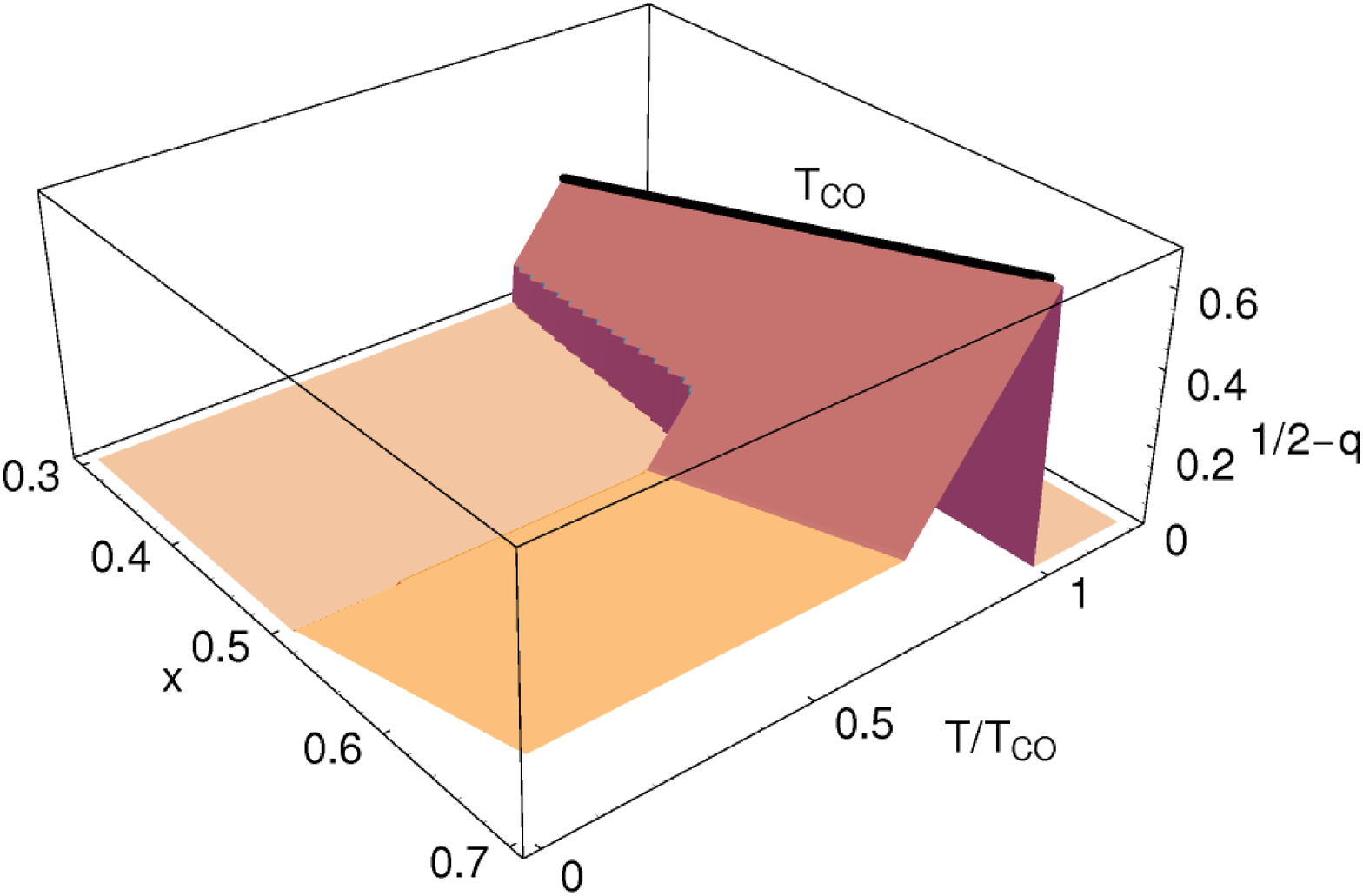, width=3.5in}
\caption{Results within Phase Modulation Approximation for a
  particular choice of parameters. These complement and clarify the schematic view
  in Fig. \ref{fig:GLphasediagram}.
Top: magnetisation in the doping-temperature plane. The low
  temperature phases are non magnetic for $x>0.5$ (corresponding to
  antiferromagnetism in experiments) and weakly ferromagnetic for
  $x<0.5$ (corresponding to the canted antiferromagnetism found in
  Pr$_{1-x}$Ca$_x$MnO$_3$ \cite{jirak85,tomioka96,yoshizawa95}).
  Above the N\'eel temperature there is a reentrant magnetisation that
  has been reported for $x \ge 0.5$.
Bottom: deviation from commensurability $1/2-q$ in the
doping-temperature plane. The surface consists of three planes that
  correspond to commensurate order $\nabla \phi=0$ for the ground
  state of the charge ordered underdoped samples, incommensurate order
  for the overdoped samples that follow $1/2-x$ as shown in the inset
  of Fig. \ref{fig:exp}, and a change of wave vector with temperature
  above the N\'eel transition as shown in the main panel of Fig.\ref{fig:exp}.
}
\label{fig:mag-eps}
\end{figure}

These features are reproduced by numerical minimisation of the coupled free
energy with appropriate values of the parameters. Figure
\ref{fig:GLphasediagram} shows the generic form of the phase diagram that
can be obtained.
%
In Fig. \ref{fig:mag-eps} we show an explicit evaluation of the
temperature- and
doping-dependence of the magnetism and (in)commensurability for parameters
chosen to approximately reproduce the experimental regimes. The
parameter range that can be used without altering the main features in
Figs. \ref{fig:GLphasediagram}, \ref{fig:mag-eps} is quite wide provided
charge modulation dominates over magnetic order.

\begin{figure}
  \epsfig{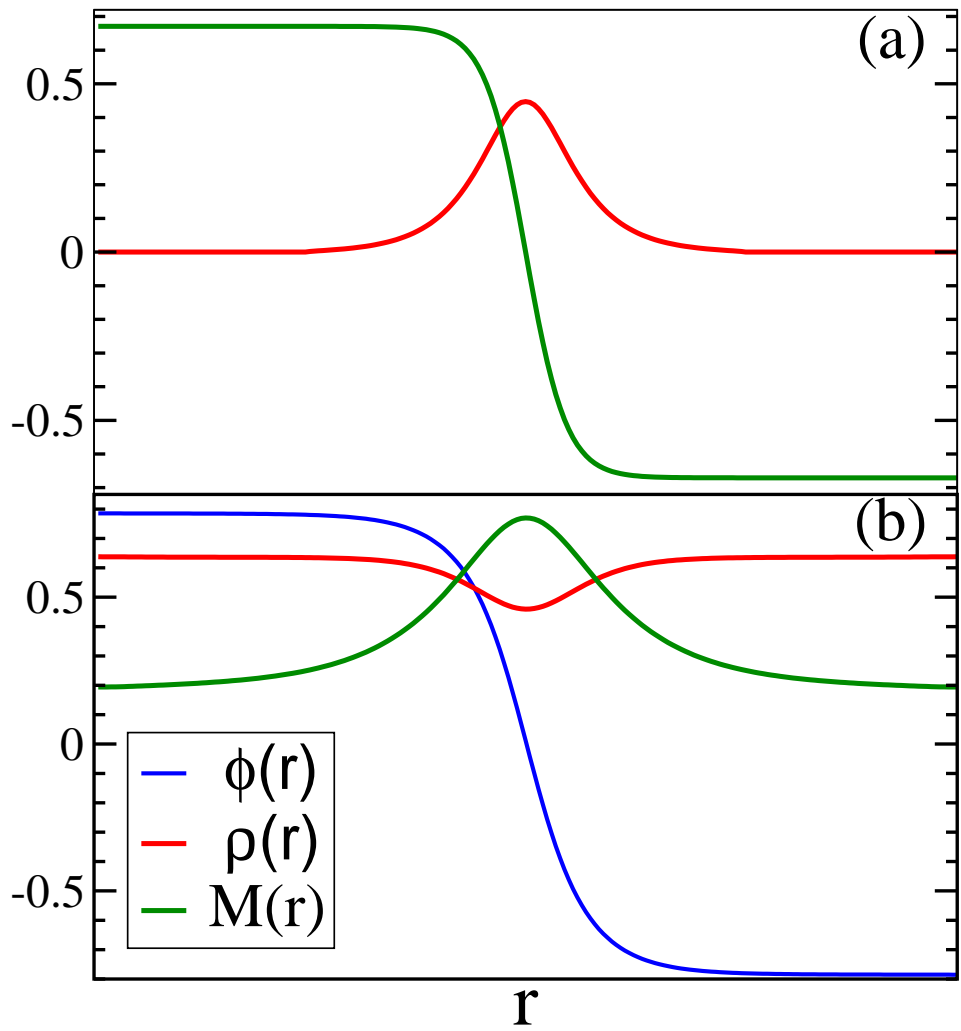}
\caption{Example solutions of the total Free Energy when all the order
  parameters are allowed to change spatially. (a) The magnetisation (green)
  decays in the centre of a magnetic domain wall leading to the appearance of
  charge modulation (red). (b) In a discommensuration, the phase of the
  charge modulation (blue) changes by $\pi/2$, the amplitude of the charge
  modulation (red) is
  suppressed, and the magnetisation (green) enhanced.
}
\label{fig:beyondPMA}
\end{figure}

Throughout much of the phase diagram shown in these figures, the order
parameters $M$, $\rho$ and $\nabla \phi$ are approximately uniform in space,
at least for the parameters we have chosen. However, close to the
commensurate-incommensurate transition, the phase modulation becomes non
uniform, and the phase gradient is built up by periodic discommensurations
where the phase advances through $2 \pi / n$. In such an inhomogeneous
state, magnetism is naturally enhanced at the boundary and the amplitude of
the charge modulation suppressed (see Fig. \ref{fig:beyondPMA}).
Another situation where
inhomogeneity is enforced is at a magnetic domain wall, where it can be
energetically
preferable to have a sharp wall stabilised by an insertion of local charge
modulation. Such a phenomenon might be responsible for the anomalously large
resistance reported for magnetic domain walls in LaCaMnO$_3$
\cite{mathur99,rzchowski}.

When interpreting experimental results in the light of the theory
presented here, a couple of issues must be kept in mind.
Firstly, the theory proposed is a mean field theory that cannot describe
strong fluctuations as its solutions are uniformly ordered or
disordered phases. However, experimentally there are some regimes, like the
incommensurate phase of Pr$_{1/2}$Ca$_{1/2}$MnO$_3$, where strong
ferromagnetic fluctuations have been measured
\cite{kajimoto01,kajimoto98}, rather than the long range magnetic order
we would predict.
Secondly, in order to keep the calculations tractable,
the phase transitions have been forced to be continuous. This
explains why the reentrant magnetism above $T_L$ shown in
Fig.~\ref{fig:GLphasediagram}
appears only when $T_C > T_{CO}$. If this theory were generalised to
include discontinuous phase transitions, this condition would
relax.
Another consequence of assuming continuous phase transitions
is that phase separation cannot be predicted.
In real systems phase separation is possible since
strain \cite{mathur-littlewood2,ahn04},
or
disorder \cite{moreo00},
can make more or less localised phases dominate within a given region.
Orbital ordering is described by a vector order parameter, thus our simple
model cannot address the complexity of different orbitally ordered phases that
have been proposed \cite{brey04}.

The Ginzburg-Landau phenomenology we propose is capable of systematizing some
puzzling data for manganites near $x = 1/2 $, but of course the propensity
for mixed and homogeneous phases is driven by the underlying physical
parameters that make the energetic cost of spatial fluctuations low. This
``electronic softness'' means that as well as spatially disordered ``phase
separation'', one finds new ordered phases which are long period arrangements
of the two competing orders.  It may indeed be that this potential for
textured electronic phases is a hallmark of electronic oxides near the Mott
transition \cite{pbl1}, seen perhaps in the co-existence of density waves and
superconductivity in the cuprates \cite{pbl2}.

{\bf Acknowledgements.}\\
The authors thank L. Brey and N. Mathur for helpful
discussions.  PBL thanks the National High Magnetic Field
Laboratory of the Los Alamos National Laboratory for hospitality.
MJC acknowledges Churchill College, University of Cambridge,  for the
award of a JRF.
This work was supported by the EPSRC and through the EPSRC Magnetic Oxide Network.

\newpage

\end{document}